\documentstyle[12pt,epsfig]{article}
\newcommand{\beq}{\begin{equation}}
\newcommand{\eeq}{\end{equation}}
\newcommand{\beqn}{\begin{eqnarray}}
\newcommand{\eeqn}{\end{eqnarray}}
\newcommand{\stackM}{\stackrel{\scriptstyle >}{{ }_{\sim}}}
\newcommand{\stackm}{\stackrel{\scriptstyle <}{{ }_{\sim}}}  
\begin{document}

\thispagestyle{empty}
\def\pubnum{444}
\def\data{July, 1998}
\begin{flushright}
{\parbox{3.5cm}{
UAB-FT-444

July, 1998

hep-ph/9807244
}}
\end{flushright}
\vspace{3cm}
\begin{center}
\begin{large}
\begin{bf}
QUANTUM SUSY SIGNATURES IN 
LOW AND HIGH ENERGY PROCESSES\\
\end{bf}
\end{large}
\vspace{1cm}
Joan SOL\`A\footnote{Invited talk at the {\it 5th International Workshop
on High Energy Physics Phenomenology} (WHEPP-5), Inter-University
Centre for Astronomy and Astrophysics (IUCAA), Pune, India ,  
January 12-26, 1998. To appear in the Proceedings.}\\

\vspace{0.25cm} 
Grup de F\'{\i}sica Te\`orica\\ 
and\\ 
Institut de F\'\i sica d'Altes Energies\\ 
\vspace{0.25cm} 
Universitat Aut\`onoma de Barcelona\\
08193 Bellaterra (Barcelona), Catalonia, Spain\\
\end{center}
\vspace{0.3cm}
\hyphenation{super-symme-tric re-nor-ma-li-za-tion}
\hyphenation{com-pe-ti-ti-ve}
\begin{center}
{\bf ABSTRACT}
\end{center}
\begin{quotation}
\noindent
In the search for phenomenological evidence of
supersymmetry through the indirect method of quantum signatures, it is useful
to seek correlations of the non-standard quantum effects in low
and high energy proceses, such as those involving on one hand the properties
of the $B$-mesons and on the other hand the physics of the top quark
and of the Higgs bosons. There are regions of the MSSM parameter space
where the potential quantum SUSY signatures in the two energy
regimes are strongly interwoven and
therefore the eventual detection of these correlated quantum effects would
strongly point towards the existence of underlying
supersymmetric dynamics.
\end{quotation} 
\newpage

\baselineskip=6.5mm  
Supersymmetry (SUSY) is perhaps the only known framework beyond the 
Standard Model (SM) which is capable of extending
non-trivially the quantum field
theoretical structure of the conventional
strong and electroweak interactions while keeping all the
necessary ingredients insuring internal consistency, such as
gauge invariance and renormalizability. A major goal of SUSY is to
produce a unified theory of all the interactions, including gravity.
At present, the simplest and most popular realization of this idea,
namely the Minimal Supersymmetric Standard Model (MSSM)\,\cite{Gunion},
is being thoroughly scrutinized by experiment and it has successfully
passed all the tests up to now.
In particular, the global fit analyses to a huge number of 
indirect precision
data within the MSSM are comparable to those in the SM\,\cite{WdeBoer}.
One could naively think that this achievement of the MSSM is basically
trivial because of the additional number of parameters involved, but
this is not necessarily true on several accounts.
First, the global fit analyses to a huge number of indirect precision
data within the MSSM are only slightly worse than in the SM 
precisely because of 
the larger number of parameters\,\cite{WdeBoer}.
Second, the healthy status of the MSSM from the point of view
of the precision measurements is {\it not} shared at all by 
any of the known alternative extensions of
the SM, like e.g. anyone of the multifarious extended 
Technicolour models.
Third, the MSSM offers
a starting point for a successful Grand Unified (GUT) framework where a
radiatively stable low-energy Higgs sector can survive.
Fourth, when embedding the MSSM into a GUT
the running of the gauge coupling constants of the 
$SU(3)_c\times SU(2)_L\times U(1)_Y$ model, starting from
the well-known initial conditions at the Fermi scale, 
turn out to match at a
proton-decay-allowed GUT scale ($\sim 10^{16}\,GeV$)
to within less than one standard 
deviation\,\footnote{Relative to the quantity
$(\alpha^{-1}_3(M_Z)-\alpha^{-1}_2(M_Z))/
(\alpha^{-1}_2(M_Z)-\alpha^{-1}_1(M_Z))=
(b_3-b_2)/(b_2-b_1)$\,\cite{Peskin} which is measurable and 
theoretically predictible, $b_i$ being the one-loop
$\beta$-function coefficients in the SUSY framework.}.
In contrast, the same running for a conventional GUT
fails to unify the gauge couplings by more than $6\,\sigma$
away of a purported unifying scale which,
to make things even worse, it is already ruled out by
the unreported observation of proton decay. 
Putting things together it is  well justified, we believe, to keep
alive all
efforts on all fronts trying to discover a supersymmetric 
particle.  Undoubtedly, the next Tevatron
run, and of course also the advent of the LHC 
and the possible
construction of an $e^+\,e^-$ supercollider (NLC),
should offer us a gold-plated
scenario for finding a direct signal of SUSY.
Then we should definitely see whether our faith has paid off.

In the meanwhile, and in view that supersymmetric particles must
be quite heavy, one naturally looks
for ``quantum signatures'' of the new physics by means of the 
indirect method of high precision measurements. 
In this talk we wish to stress the
possibility of seeing large virtual effects of SUSY through the 
correlation of the quantum effects in the low and high energy
domains. Thus on the one hand at high energies we have above all the
physics of the top quark and the potential existence of the Higgs
bosons. Indeed, one
expects that a first hint of Higgs activity, if ever, should appear in
concomitance with the detailed studies of top quark phenomenology.
On the other hand, in the low energy domain, $B$-meson physics
could be a serious competitor to high energy processes in the search
for extensions of the SM.   
Not in vain the restrictions 
placed by e.g. radiative $B^0$ decays
$\bar{B}^0\rightarrow X_s\,\gamma$, i.e.
\beq
b\rightarrow s\,\gamma\,,
\label{eq:bsgamma}
\eeq
on the global fit analyses\,\cite{WdeBoer} to indirect precision
electroweak data have played a significant role. 
Although at present the SM prediction on the decay (\ref{eq:bsgamma})
is only a little bit higher than the CLEO results\,\cite{CLEO}
and the discrepancy seems no longer statistically
significant, the precise knowledge of this observable is particularly useful
to restrain new forms of physics beyond the SM. In fact,
in the absence of SUSY, the CLEO data alone\,\cite{CLEO} on the decay
(\ref{eq:bsgamma}) preclude general Type II two-Higgs-doublet models
($2$HDM's)\,\cite{Hunter} involving typical charged  Higgs masses 
$M_{H^\pm}\stackm 250\,GeV$ \,\cite{Ciuchini}, 
thus barring the possibility
of the non-standard top quark decay $t\rightarrow H^+\, b$. Still we
should point out that the
inclusion of the ALEPH data\,\cite{ALEPH}, with a larger central value,
 would not completely close this channel. 
In any case a bound on the charged Higgs mass is there and stems from
the fact that charged Higgs
bosons of ${\cal O}(100)\,GeV$ interfere constructively with the
SM amplitude of the decay (\ref{eq:bsgamma}) and render
a final value of $BR(b\rightarrow s\,\gamma)$ exceedingly high. 
This situation
can be remedied in the MSSM, where there may be
a compensating contribution from
relatively light charginos
and stops which tend to cancel the Higgs effects\,\footnote{The recent calculation
of the QCD NLO effects in the MSSM within the ``heavy squark-gluino effective theory''
significantly reduces the previus (LO) values of the lightest
chargino-stop masses by which the SUSY contribution produces a large destructive
interference with the charged-Higgs boson contribution\,\cite{Ciuchini2}. However,
the final values of the sparticle masses remain in the few hundred $GeV$ range. }. 
Thus, we see that from $B$-meson physics a door is still well open 
within the MSSM context for the non-SM top quark decay
$t\rightarrow H^+\,b$\,\cite{CGGJS} which could compete with
the SM mode $t\rightarrow W^+\,b$. Moreover, charged Higgs exchange may 
also play a role in semileptonic
$B$-meson decays, especially into $\tau$-leptons,
\begin{equation}
b\rightarrow c\,\tau^-\,\bar{\nu}_{\tau}\,,
\label{eq:bctau}
\end{equation}
where the $2HDM$ effects could significantly modify the SM 
expectations\,\cite{Grossman}.
Remarkably enough, the MSSM quantum corrections
to $B$-meson decay processes (\ref{eq:bsgamma})-(\ref{eq:bctau})
on the one hand, and to the top quark decays
\beqn
&&t\rightarrow W^+\,b\,,\nonumber\\
&&t\rightarrow H^+\,b\,,
\label{eq:tWH}
\eeqn
and the $\tau$-lepton charged Higgs decay
\beq
H^+\rightarrow \tau^+\,\nu_{\tau}
\label{eq:Htau}
\eeq
on the other, can be correlated in certain regions of parameter
space; e.g. they could
be both maximized in the same domain.
This would be a dramatic example of
low and high energy correlations that 
could be a clue to the discovery of ``virtual SUSY''. 

Ultimately, the potential connection
between the quantum effects in the two energy regimes stems from the
role played by the Yukawa sector.
Within the SM the physics of the top quark is intimately connected
with that of the Higgs sector through the Yukawa couplings. 
However, if this is true in the SM, the more it should be
in the MSSM where both the Higgs and the top quark sectors are 
virtually ``doubled'' with respect to the SM.
As a consequence, the Yukawa coupling sector is richer in the
supersymmetric model than in the standard one. This
could greatly modify the phenomenology already at the level of
quantum effects on electroweak observables.
As a matter of fact in the MSSM the 
bottom-quark Yukawa coupling may counterbalance the 
smallness of the bottom
mass at the expense of a large value of $\tan\beta$ --the ratio
$v_2/v_1$ of the vacuum
expectation values of the two Higgs doublets-- the upshot being that
the top-quark and bottom-quark Yukawa couplings in the superpotential
\beq
h_t={g\,m_t\over \sqrt{2}\,M_W\,\sin{\beta}}\;\;\;\;\;,
\;\;\;\;\; h_b={g\,m_b\over \sqrt{2}\,M_W\,\cos{\beta}}\,,
\label{eq:Yukawas} 
\eeq
can be of the same order of magnitude, perhaps even showing up 
in ``inverse'' hierarchy: $h_t<h_b$ for $\tan\beta> m_t/m_b$. 
Notice that due to the perturbative bound
$\tan\beta\stackrel{\scriptstyle <}{{ }_{\sim}}60$ one never reaches
a situation where $h_t<<h_b$.
In a sense $h_t\sim h_b$  could be judged as a natural 
relation in the MSSM.
Thus from the practical point of view, one should not dismiss the
possibility that the bottom-quark Yukawa coupling could play a
momentous role in the phenomenology of $B$-meson decays and top quark
and MSSM Higgs boson decay and production, to the extend of
drastically changing standard expectations on the observables
associated to them, such as decay widths and 
cross-sections.

\begin{figure}
\centering
\mbox{\epsfig{file=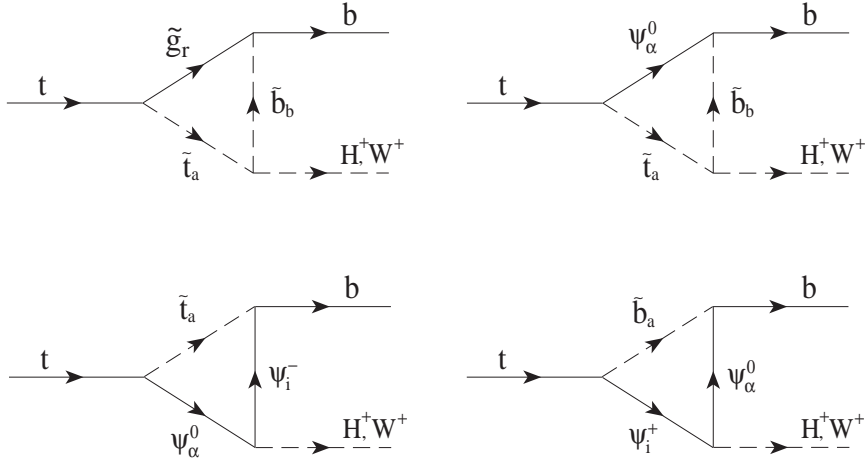,width=11.5cm}}
\caption
{SUSY-QCD and SUSY-EW one-loop vertices for 
$t\rightarrow H^+\,b$ and $t\rightarrow W^+\,b$.
The $\Psi$'s are chargino and neutralino
unphysical mass-eigenstates related to the  physical mass-eigenstates
($\chi$'s) as explained in Ref.\,\protect{\cite{Guasch1}}.}
\end{figure}
Needless to say, direct top quark decays into SUSY 
particles are in principle possible as well,
both as 2-body and 3-body final states.
Among the 2-body channels carrying an explicit
SUSY signature, the following decays stand out:
\beqn
{\rm i}) &&t\rightarrow \tilde{t}_i\,\chi^0_{\alpha},\nonumber\\
{\rm ii}) &&t\rightarrow \tilde{b}_i\,\chi^+_{\alpha},\nonumber\\
{\rm iii}) &&t\rightarrow \tilde{t}_i\,\tilde{g}\,. 
\label{eq:decays}
\eeqn
Therein, $\tilde{t}_{i}$, $\tilde{b}_{i}$, $\chi^+_i$,
$\chi^0_{\alpha}$, $\tilde{g}$ ($i=1,2;\,\alpha=1,2,...,4$)
denote stop, sbottom, chargino, neutralino and gluino sparticles,
respectively. (Decay iii assumes that a light gluino window is still
open, though it is almost ruled out at present.)
Also quite a few three-body top quark decays are possible
in the MSSM and have been 
studied in detail\,\cite{Guasch1}.
Here, however, we assume that sparticles (except perhaps the
stop) are heavy enough that the above
direct top SUSY decays ii)-iii) are forbidden.  We 
are thus mainly concerned with
the SM decay and the charged Higgs decay of the top quark.
In fact, while the decays (\ref{eq:tWH}) are not necessarily MSSM 
processes, we wish to study  whether hints of SUSY 
can be recognized out of them from pure supersymmetric quantum effects.

In the following we will display some numerical results
accounting for the potential
supersymmetric quantum corrections underlying the low and
high energy decays (\ref{eq:bctau})-(\ref{eq:Htau}) and comment
on the possibility of seeing correlated SUSY 
quantum signatures in low and high energy processes. 
We perform the analysis of quantum effects in the
on-shell $G_F$-scheme, which is characterized by the set of inputs
$(G_F, M_W,M_Z,m_f,M_{SUSY},...)$, and we use the process 
(\ref{eq:Htau}) to define (and renormalize) $\tan\beta$\,\cite{CGGJS}.
The supersymmetric
strong (SUSY-QCD) and the supersymmetric electroweak (SUSY-EW) 
one-loop vertex diagrams for the charged Higgs decay of the top quark,
$t\rightarrow H^+\,b$, are displayed in Fig.\,1.
The bottom mass corrections in Fig.\,2, as well as the corresponding
$\tau$-lepton mass counterterms associated to our definition of $\tan\beta$,
turn out to be very important.
The corresponding diagrams for the standard top quark decay,
$t\rightarrow W^+\,b$, are obtained by just replacing $H^+$
with $W^+$ in Fig.\,1, but
the mass counterterms
of Fig.\,2 play no role since in this case there is no need to renormalize
$\tan\beta$.  Both decay processes (\ref{eq:tWH})
are well understood in the MSSM\,\cite{CGGJS,GJSH}. 

Here we shall mainly report on
$t\rightarrow H^+\,b$ because the standard top quark decay 
gives a small yield.
In fact, in the on-shell $G_F$-scheme the SUSY quantum corrections to
the standard decay $t\rightarrow W^+\,b$
are negative and of the order of a few per cent (except in some
unlikely cases). Therefore, they
approximately cancel out
against the positive ``SM'' electroweak contribution\,\footnote{Within the MSSM
 context, the ``SM'' electroweak contribution is defined
to be the one obtained after decoupling the $R$-odd effects and letting the
mass of the CP-odd Higgs boson of the MSSM go to infinity\,\cite{GJSH}.}
(of the same order of magnitude) 
leaving the ordinary QCD effects  
($\simeq -10\%$) as the net MSSM corrections (Cf. Fig.\,3a).
Hence no significant
imprint of underlying SUSY dynamics is left behind 
$\Gamma(t\rightarrow W^+\,b)$ and in this way 
we are naturally led to a closer examination of the charged Higgs decay of the
top quark. To be sure,
$t\rightarrow H^+\,b$ has been object of many 
studies in the past (Cf. \cite{CGGJS} and references therein),
mainly within the context of
$2$HDM's, and it is being thoroughly searched in recent analyses at the 
Tevatron\,\cite{CDF}. Notwithstanding, no systematic treatment of the
MSSM quantum effects existed in the literature until very
recently\,\cite{CGGJS}.

\begin{figure}
\centering 
\mbox{\epsfig{file=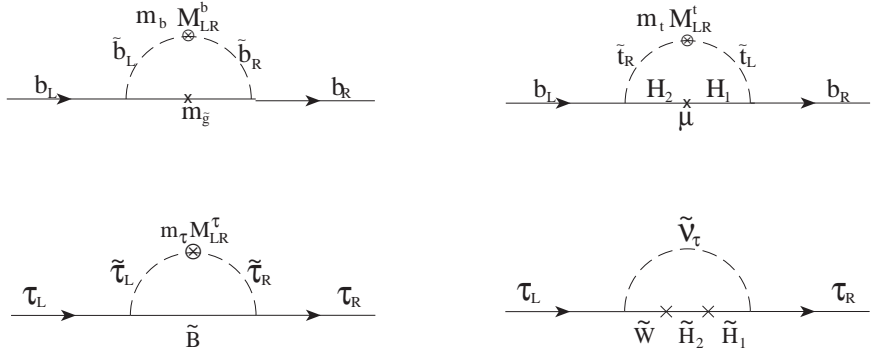,width=11.5cm}}
\caption
{Relevant bottom and $\tau$-lepton mass counterterms for $t\rightarrow H^+\,b$
and $H^+\rightarrow\tau^+\,\nu_{\tau}$. 
Here $M_{LR}^{b,\tau}=A_{b,\tau}-\mu\,\tan\beta$,
and $M_{LR}^t=A_t-\mu\,\cot\beta$, with  $\mu$  the higgsino
mass parameter and the $A$'s the corresponding trilinear
soft SUSY-breaking couplings.}
\end{figure}
To appraise the relative importance of the 
various types of MSSM effects on $\Gamma (t\rightarrow H^+\,b)$, 
in Figs.\,3b-3d we provide plots for the correction 
to the partial width as a function of $\tan\beta$, 
the lightest sbottom mass ($m_{\tilde{b}_1}$) and the gluino
mass ($m_{\tilde{g}}$) respectively, reflecting also the
various individual contributions.
Specifically, we show in these figures:
\begin{itemize}
\item{(i)} The supersymmetric electroweak
contribution from genuine ($R$-odd) sparticles (denoted $\delta_{\rm SUSY-EW}$), i.e. 
from sfermions (squarks and sleptons), charginos and neutralinos;
\item{(ii)} The electroweak contribution from non-supersymmetric ($R$-even) 
particles ($\delta_{EW}$). It is composed of two distinct types
of effects, namely, those from Higgs 
and Goldstone bosons (collectively called
``Higgs'' contribution, and denoted $\delta_{\rm Higgs}$) plus
the leading  SM effects
from conventional fermions ($\delta_{\rm SM}$):
\beq
\delta_{EW}=\delta_{\rm Higgs}+\delta_{\rm SM}\,;
\eeq
The remaining non-supersymmetric electroweak effects
are subleading and are neglected.
\item{(iii)} The strong supersymmetric contribution
(denoted by $\delta_{\rm SUSY-QCD}$) from squarks and gluinos; 
\item{(iv)}
The strong contribution from conventional quarks
and gluons (labelled $\delta_{\rm QCD}$);
 and
\item{(v)}
The total MSSM contribution, $\delta_{\rm MSSM}$,
namely, the net sum of all the previous contributions:
\beq
\delta_{\rm MSSM}=\delta_{\rm SUSY-EW}+\delta_{EW}+
\delta_{\rm SUSY-QCD}+\delta_{\rm QCD}.
\label{eq:individ}
\eeq
\end{itemize}
We remark in Fig.\,3d the local maximum in the gluino contribution 
(around $m_{\tilde{g}}=500\,GeV$) which entails
a net MSSM correction $\delta_{\rm MSSM}\simeq 20\%$ and $75\%$ 
at $\tan\beta=30$ and $60$ respectively. This decay
is manifestly much more sensitive to heavy (not light!) gluinos
($\delta_{\rm SUSY-QCD}\simeq 0$ for $m_{\tilde{g}}=0$).
Only for superheavy gluinos $m_{\tilde{g}}>>1\,TeV$ the effect
eventually decouples.

We may easily understand the reason why $t\rightarrow W^+\,b$ cannot
generate comparably large quantum SUSY signatures as $t\rightarrow H^+\,b$.  
The counterterm configuration associated to vertices involving
gauge bosons and conventional fermions does {\it not}
involve the term\,\cite{SO10,CGGJS}
$\delta m_b/m_b$ (Cf. Fig.2) --which grows (approx.) linearly with
$\tan\beta$ -- so that one
cannot expect similar enhancements. Therefore, the SUSY-QCD
corrections to $t\rightarrow W^+\,b$ 
are not foreseen to be particularly significant in this case,
but just of order $\alpha_s(m_t)/4\,\pi$.
This is borne out by the numerical analysis in Fig.3a.
The only hope for gauge boson 
interactions with top and bottom quarks to develop sizeable radiative
corrections was to appeal to large non-oblique
corrections triggered by the Yukawa terms (\ref{eq:Yukawas}).
However, even in this circumstance the results are rather disappointing.
For, at large $\tan\beta\geq m_t/m_b$ the
bottom quark Yukawa coupling (the only relevant one in these conditions)
gives a contribution of order\,\cite{GJSH} 
\beq
{\alpha_W\over 4\,\pi}\,{m_b^2\,\tan^2\beta\over M_W^2}\stackM 
{\alpha_W\over 4\,\pi}\,{m_t^2\over M_W^2}\,,
\label{eq:Yb}
\eeq
which is numerically very close to the strong contribution $\alpha_s (m_t)/4\,\pi$.
In contrast, the typical SUSY-QCD effect on the decay
$t\rightarrow H^+\,b$ (the leading type of effect for this mode) in the limit
where there is no large hierarchy
between the sparticle masses is of order\,\cite{CGGJS}
\beq
C_F\,{\alpha_s\over 4\,\pi}\,\,\tan\beta\,,
\label{eq:dmb2}
\eeq
where $C_F=4/3$ is the eigenvalue of the quadratic Casimir
operator of $SU(3)$ in the fundamental representation. 
The ratio between (\ref{eq:dmb2})
and (\ref{eq:Yb}) reads
\beq
C_F\,\left({\alpha_s\over\alpha_W}\right)\,
\,\left({M_W\over m_t}\right)^2\,
\tan\beta= {\cal O}(1)\,\tan\beta\,.
\label{eq:rat}
\eeq
Consequently, in the entire high $\tan\beta$ regime the SUSY-QCD effects
on the $t\,b\,H^{\pm}$-vertex are expected to
be a factor of order $\tan\beta$
larger than the SUSY-QCD and 
Yukawa coupling effects on $t\,b\,W^{\pm}$\,\footnote{Of course, inclusion of the
leading (Yukawa type) electroweak effects on $t\rightarrow H^+\,b$ further enhances
the ratio (\ref{eq:rat}). The ``exact'' numerical result is contained in Fig.\,3b.}.
\begin{figure}
\centering
        \begin{tabular}{cc}
          \mbox{\epsfig{file=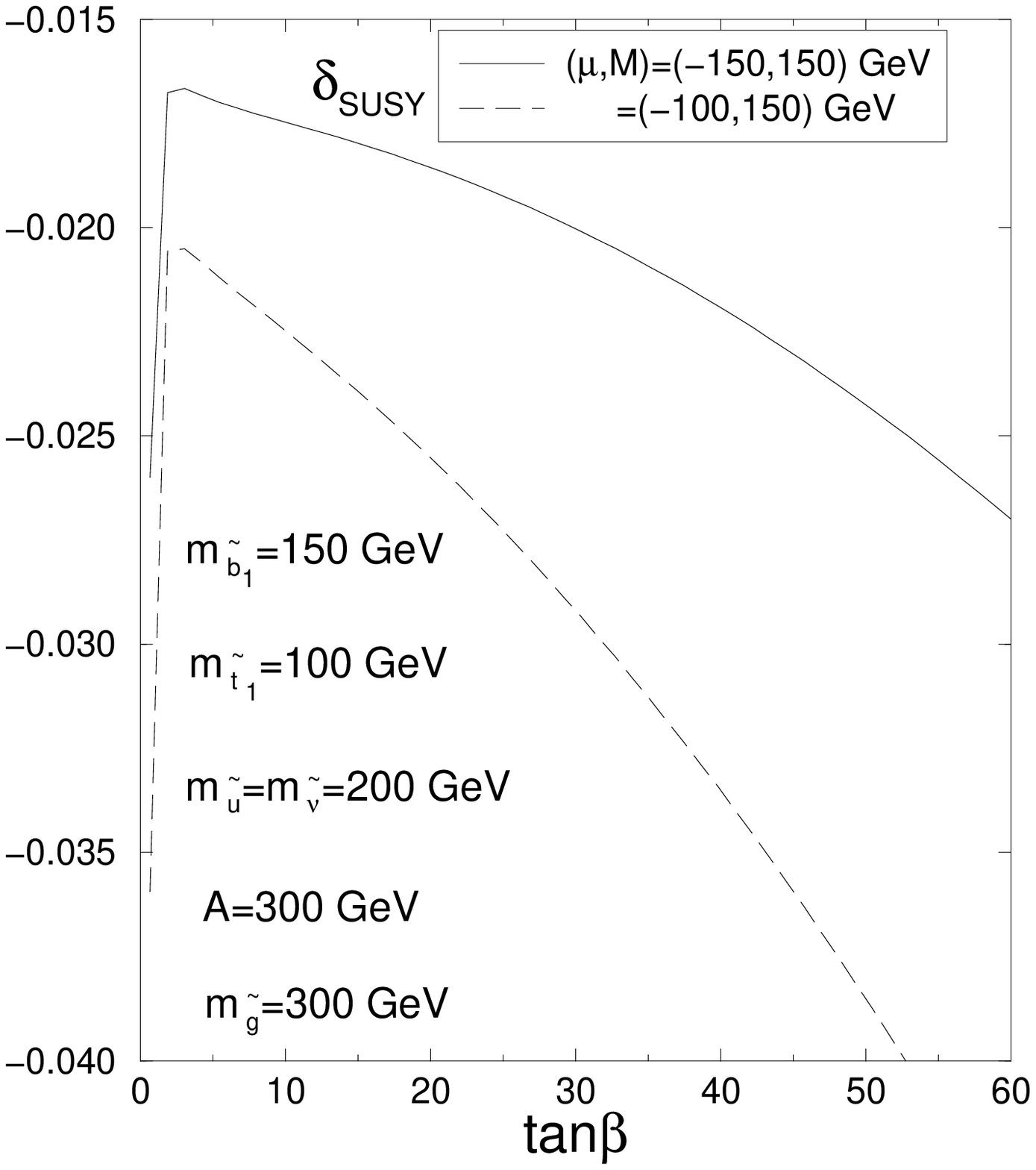,width=5.5cm}}&
          \mbox{\epsfig{file=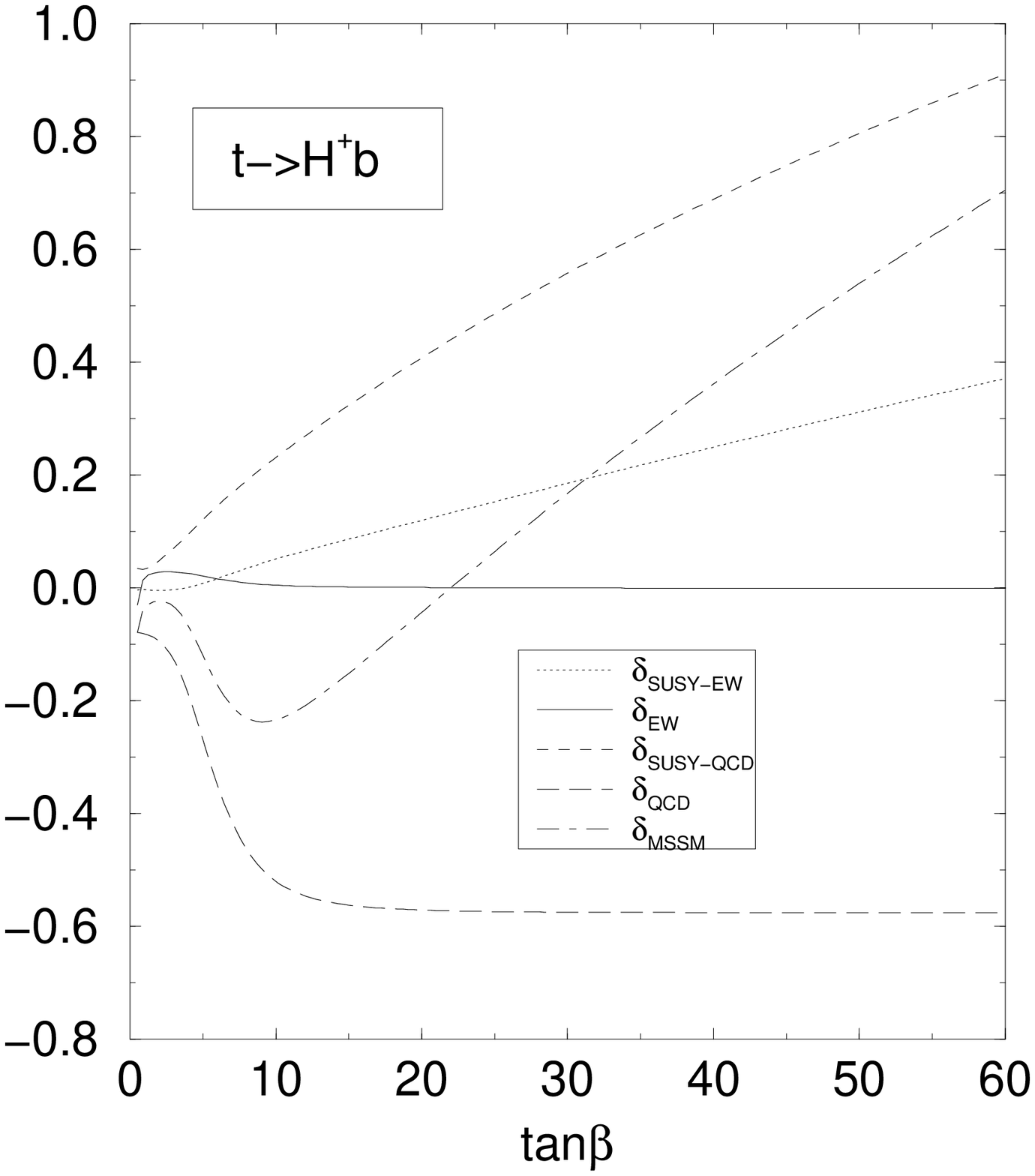,width=5.5cm}} \\(a)&(b)\\[0.5cm]
          \mbox{\epsfig{file=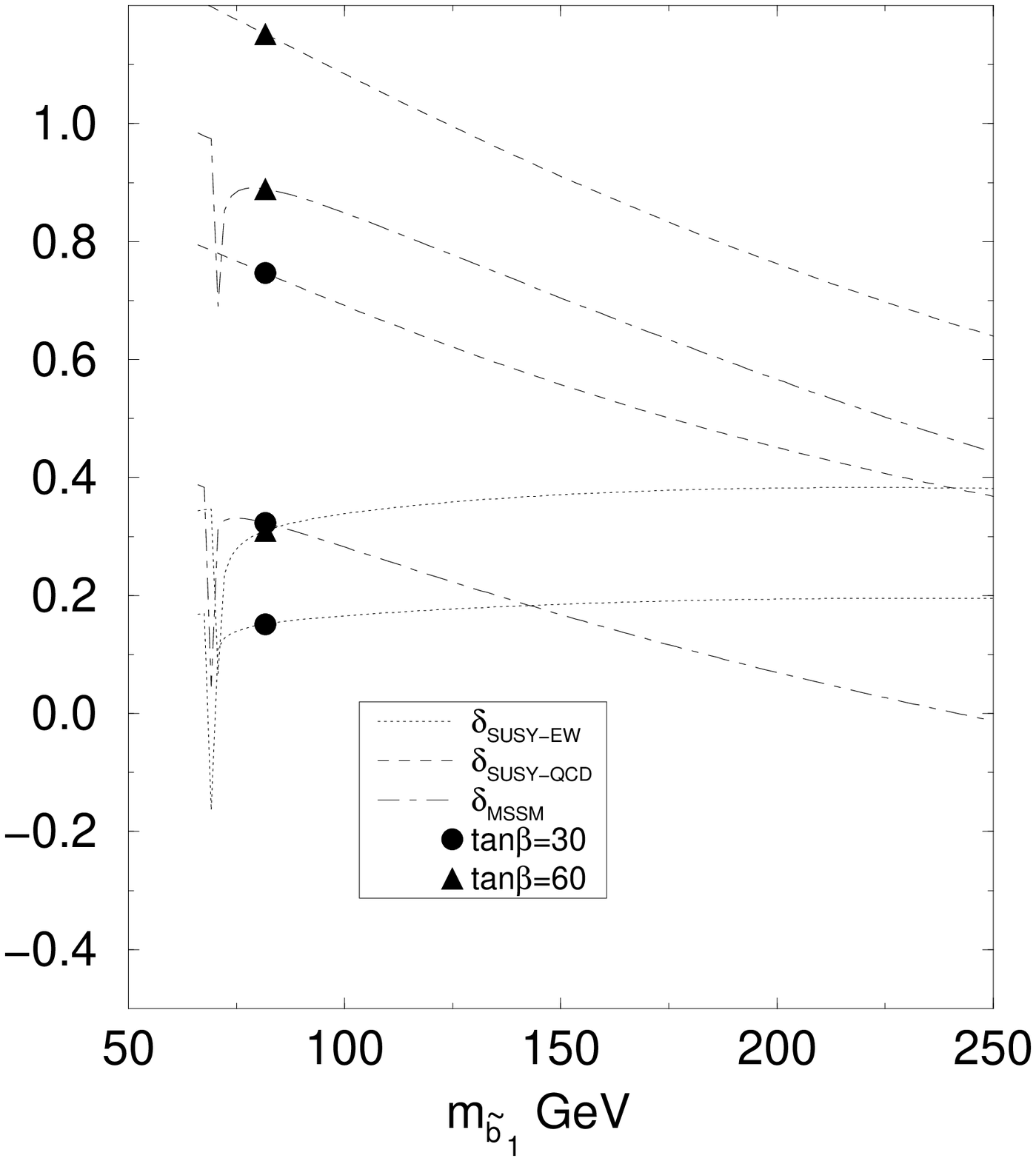,width=5.5cm}}&
          \mbox{\epsfig{file=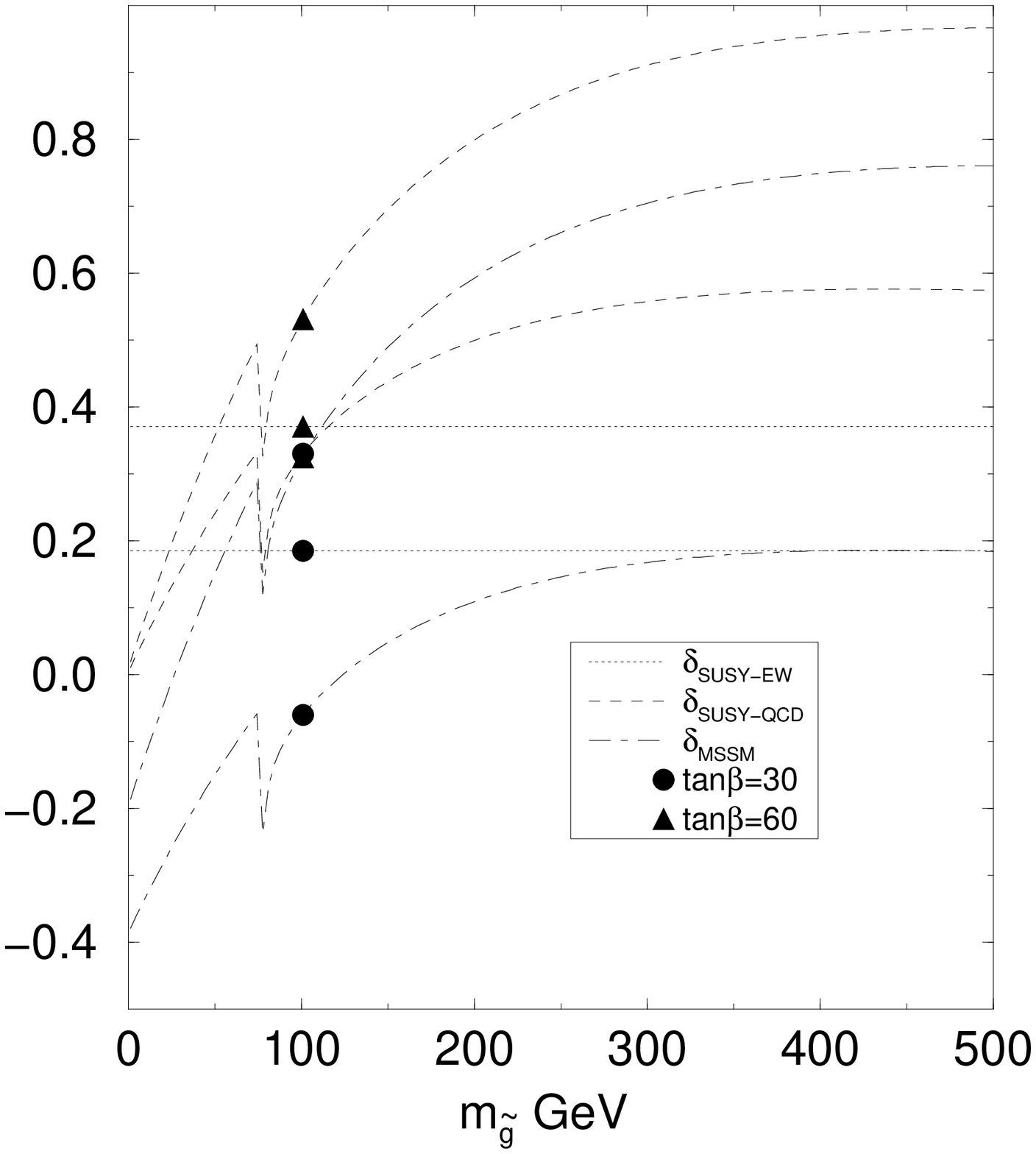,width=5.5cm}} \\(c)&(d)
        \end{tabular}
\caption
{{\bf (a)} The total (electroweak and strong) SUSY correction to
$\Gamma (t\rightarrow W^+\,b)$ for given sets of parameters. Notation
as in Ref.\,\protect{\cite{CGGJS}}.
{\bf (b)} The SUSY-EW, SUSY-QCD, standard QCD
and full MSSM corrections to	
$\Gamma (t\rightarrow H^+\,b)$ as a function of
$\tan\beta$; {\bf (c)} As in (b), but as a function of
the lightest sbottom mass; 
{\bf (d)} As in (b), but as a function of
the gluino mass.}
\end{figure}

As a practical application of the previous analysis of SUSY quantum
effects on $t\rightarrow H^+\,b$, let us consider the implications 
derived from the non-observation of an excess of
$\tau$-events from $H^\pm$ decays (\ref{eq:Htau}) at the Tevatron\,\cite{CDF}.
Our definition 
of $\tan\beta$ from the vertex associated to the decay (\ref{eq:Htau})
allows to renormalize the
$t\,b\,H^{\pm}$-vertex in perhaps the most convenient way to deal with 
$t\rightarrow H^+\,b$.  Indeed, from the practical point of view, we should
recall the excellent methods for $\tau$-identification 
developed by the Tevatron
collaborations and recently used by CDF to study the 
permitted region in
the $(\tan\beta,M_H)$-plane\,\cite{CDF}.
However, we wish to show that this analysis may undergo dramatic 
changes when we incorporate the MSSM
quantum effects\,\cite{Guasch4}. Although CDF utilizes
inclusive $\tau$-lepton tagging, for our purposes it will
suffice to focus on the exclusive final state
$(l,\tau)$, with $l$ a light
lepton, as a means for detecting 
an excess of $\tau$-events\,\cite{DPRoy}.
To be precise, we are interested in the
$t\,\bar{t}$ cross-section leading to the decay sequences  
$t\,\bar{t}\rightarrow H^+\,b,W^-\,\bar{b}$ and 
$H^+\rightarrow \tau^+\,\nu_{\tau}$, $W^-\rightarrow l\,\bar{\nu}_l$, 
and {\it vice versa}. 
The relevant quantity can be
easily derived from the measured value
of the canonical cross-section $\sigma_{t\bar{t}}$ for the standard
channel $t\rightarrow b\,l\,\nu_l$, $\bar{t}\rightarrow b\,q\,q'$, after
inserting appropriate branching fractions, namely\,\cite{Guasch4} 
\beq
\sigma_{l\tau}=\left[\frac4{81}\,\epsilon_1+\frac49\,
{\Gamma (t\rightarrow H\,b)\over
\Gamma (t\rightarrow W\,b)}\,\epsilon_2\right]\,\sigma_{t\bar{t}}\,.
\label{eq:bfrac}
\eeq 
The first term in the bracket comes from the SM top quark decay, and 
for the second term we assume (at high $\tan\beta$) $100\%$ branching 
fraction of $H^+$ into $\tau$-lepton, as explained before. Finally,
$\epsilon_i$ are detector efficiency factors. 
Thus, in most of the phase space available for top decay
the bulk of the cross-section (\ref{eq:bfrac}) is provided by
the contribution of $\Gamma(t\rightarrow H^+\,b)$. Consequently, the 
observable (\ref{eq:bfrac})
should be highly sensitive to MSSM quantum effects.

\begin{figure}
\centering
        \begin{tabular}{cc}
          \mbox{\epsfig{file=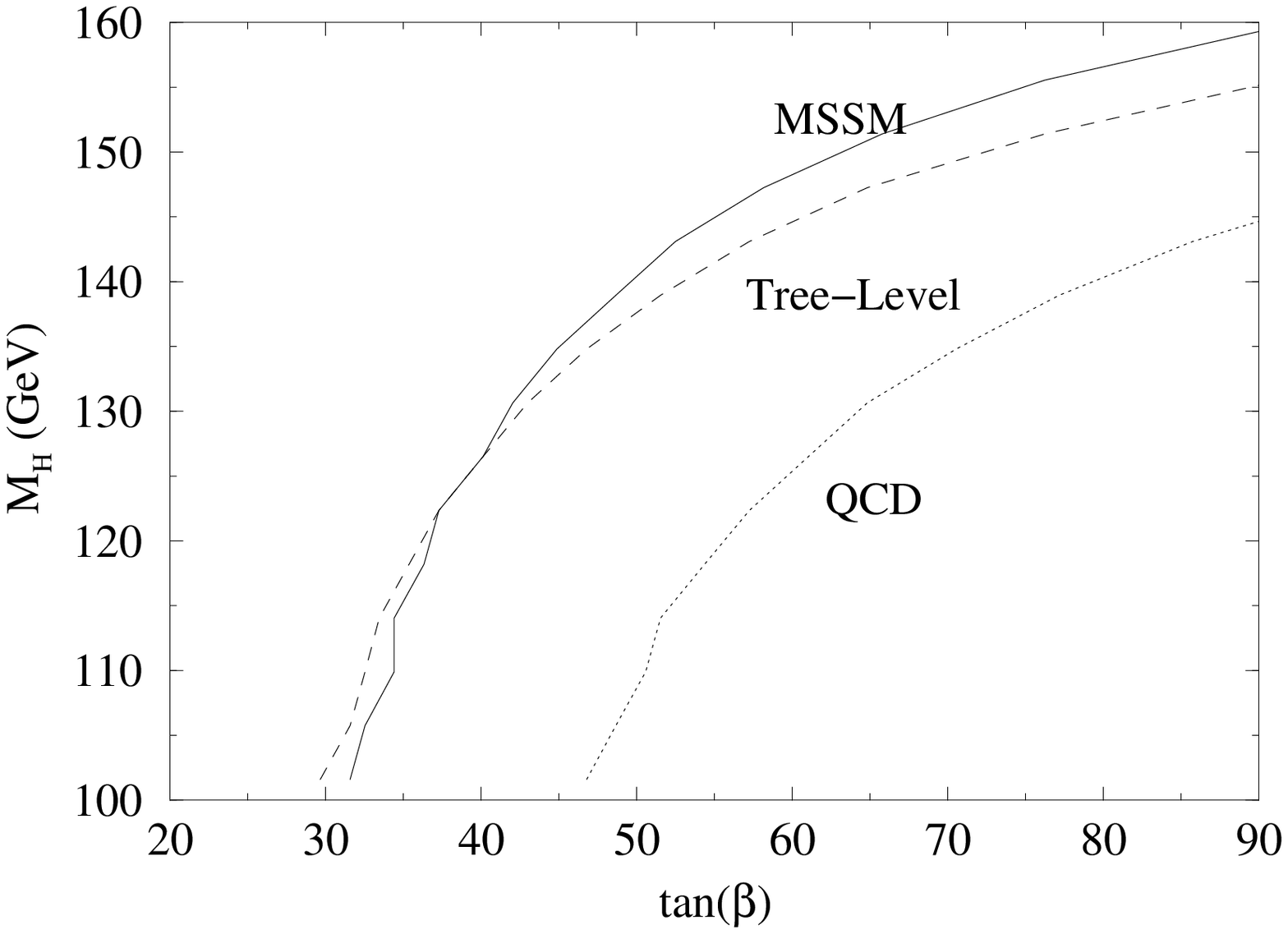,width=5.5cm}}&
          \mbox{\epsfig{file=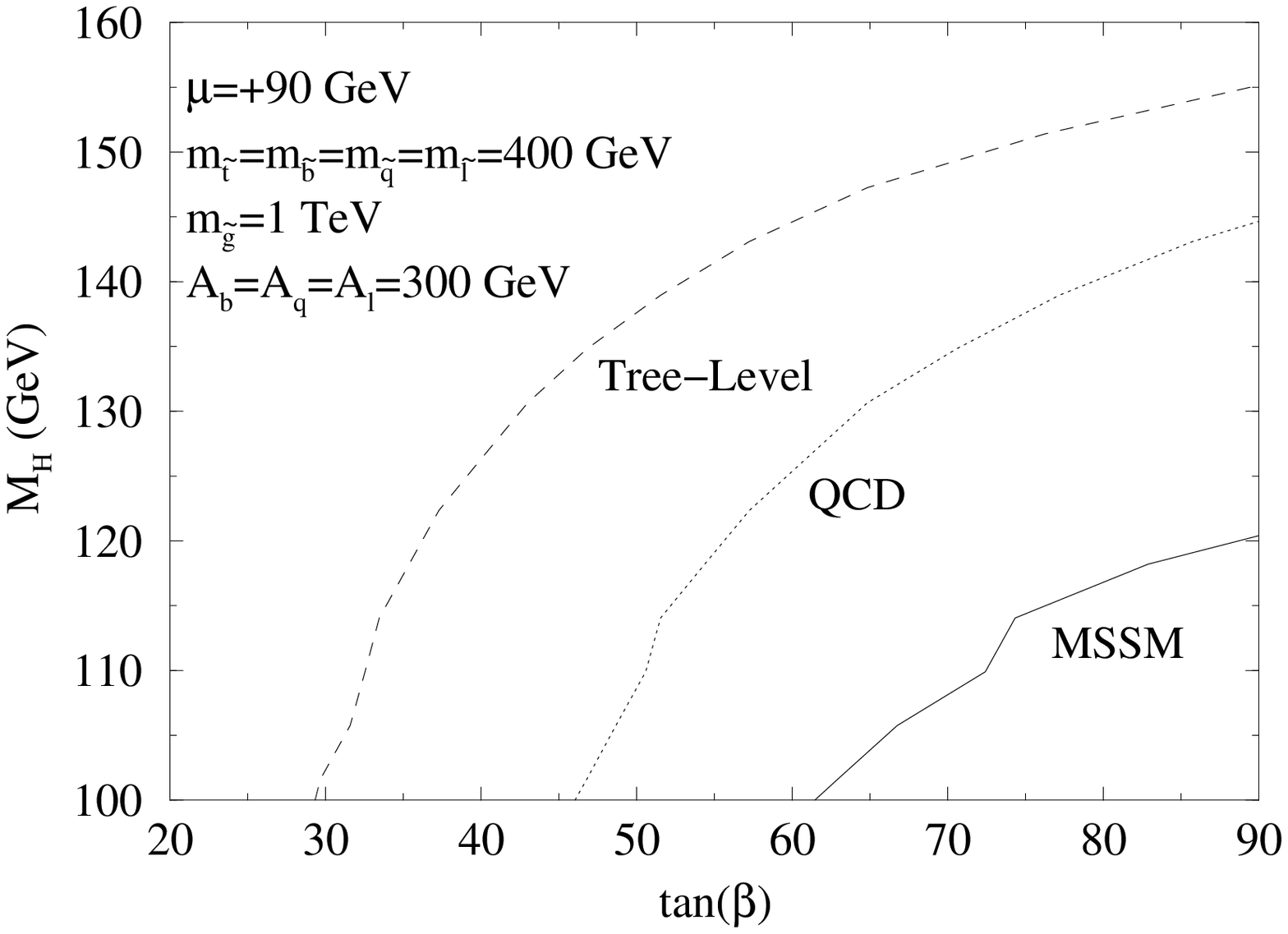,width=5.5cm}} \\(a)&(b)
        \end{tabular}
\caption
{{\bf (a)} The $95\%$ C.L. exclusion plot in the 
$(\tan\beta, M_{H^\pm})$-plane
for $\mu=-90\,GeV$ and remaining parameters similar to Fig.\,3.
Shown are the tree-level (dashed), QCD-corrected
(dotted) and fully 
MSSM-corrected (continuous) contour lines.
The excluded region in each case is the one lying below the curve;
{\bf (b)} As in (a), but for a $\mu>0$ scenario characterized by a 
heavier SUSY spectrum which makes the analysis compatible
with perturbation theory.}
\end{figure}


In Figs.\,4a and 4b we derive the ($95\%$ C.L.) excluded regions for
$\mu<0$ and $\mu>0$, respectively. (In the $\mu>0$ case we choose a
heavier SUSY spectrum in order that the correction remains
perturbative\,\cite{Guasch4}.)
We point out that the numerical results in Figs.\,4a-4b include
the restrictions on the MSSM parameter space placed by the
radiative $B$-meson decay, eq.(\ref{eq:bsgamma})\,\footnote{For
detailed plots of the MSSM domain allowed by $b\rightarrow s\,\gamma$,
see e.g. Ref.\,\cite{CGGJS2}.}.
From inspection of these figures
it can hardly be overemphasized that the MSSM quantum effects on the
CDF analysis\,\cite{CDF}
can be dramatic. In particular, while for $\mu<0$ the MSSM-corrected
curve is significantly more restrictive than 
the QCD-corrected one, 
for $\mu>0$ the bound essentially disappears from the perturbative
region ($\tan\beta\stackm 60$). 

We point out the recent work of Ref.\,\cite{Guchait1} claiming 
that the bound on the $(\tan\beta, M_H)$-plane substantially improves
using  Tevatron data in
the $b\bar{b}\tau^+\tau^-$ channel. Nonetheless this calculation was 
performed by considering only conventional QCD corrections 
and hence it could still undergo significant
MSSM radiative corrections. The potentially large
effects not included in that paper
stem from the production mechanism of the CP-odd Higgs boson $A^0$
(through $b\bar{b}$-fusion) before it decays into $\tau^+\tau^-$ pairs. 
Indeed, the $b\bar{b}\,A^0$ vertex is known\,\cite{Coarasa} to 
develop important MSSM corrections
in the relevant regions of the $(\tan\beta, M_H)$-plane purportedly
``excluded'' by the tree-level analysis of Ref.\cite{Guchait1}.
Therefore, a detailed
re-examination of the excluded region at the quantum level is in order
within the context of the MSSM before jumping into conclusions.

\begin{figure}
\centering
\begin{tabular}{c}
\mbox{\epsfig{file=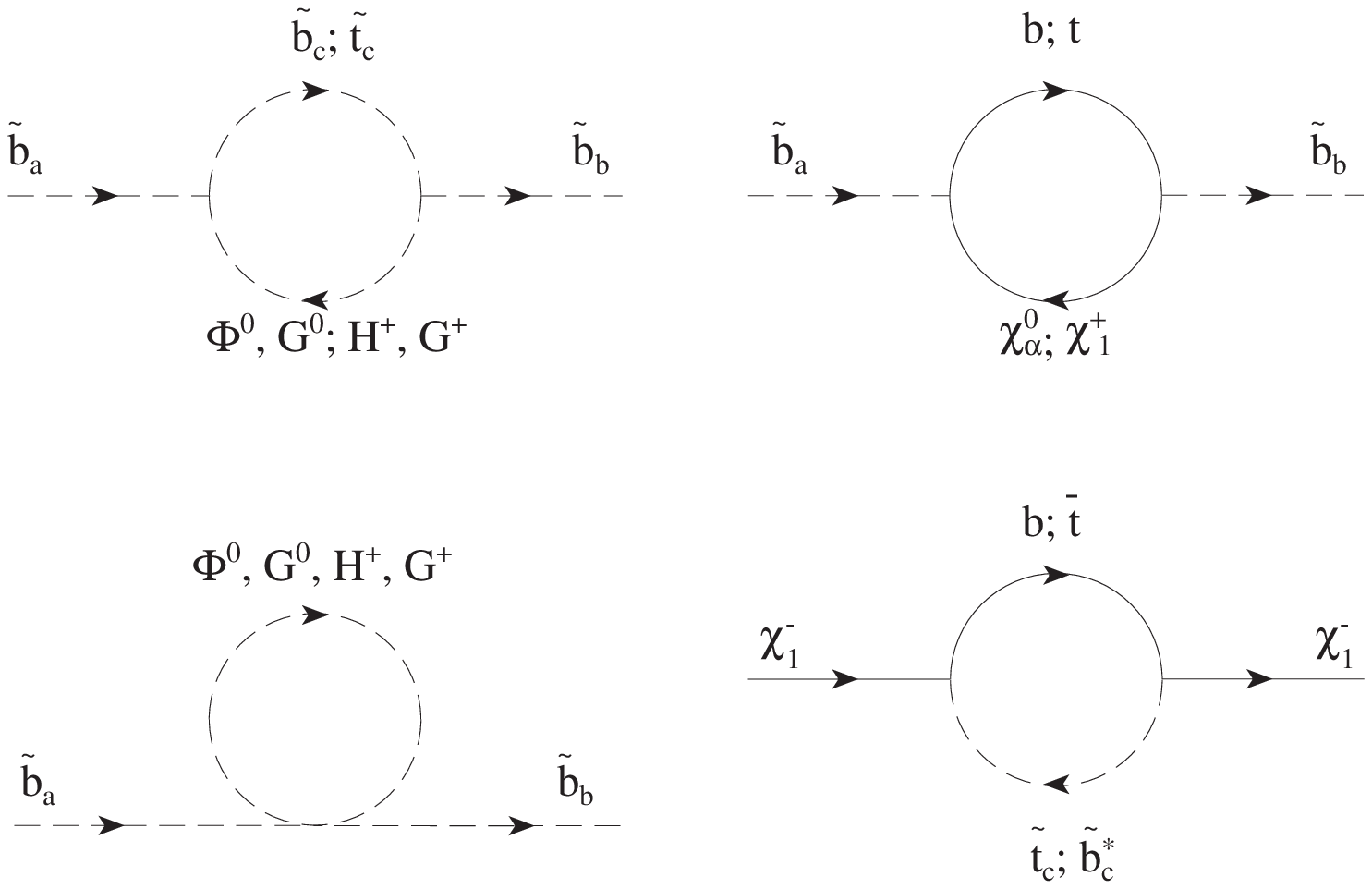,width=11cm} }\\
\mbox{\epsfig{file=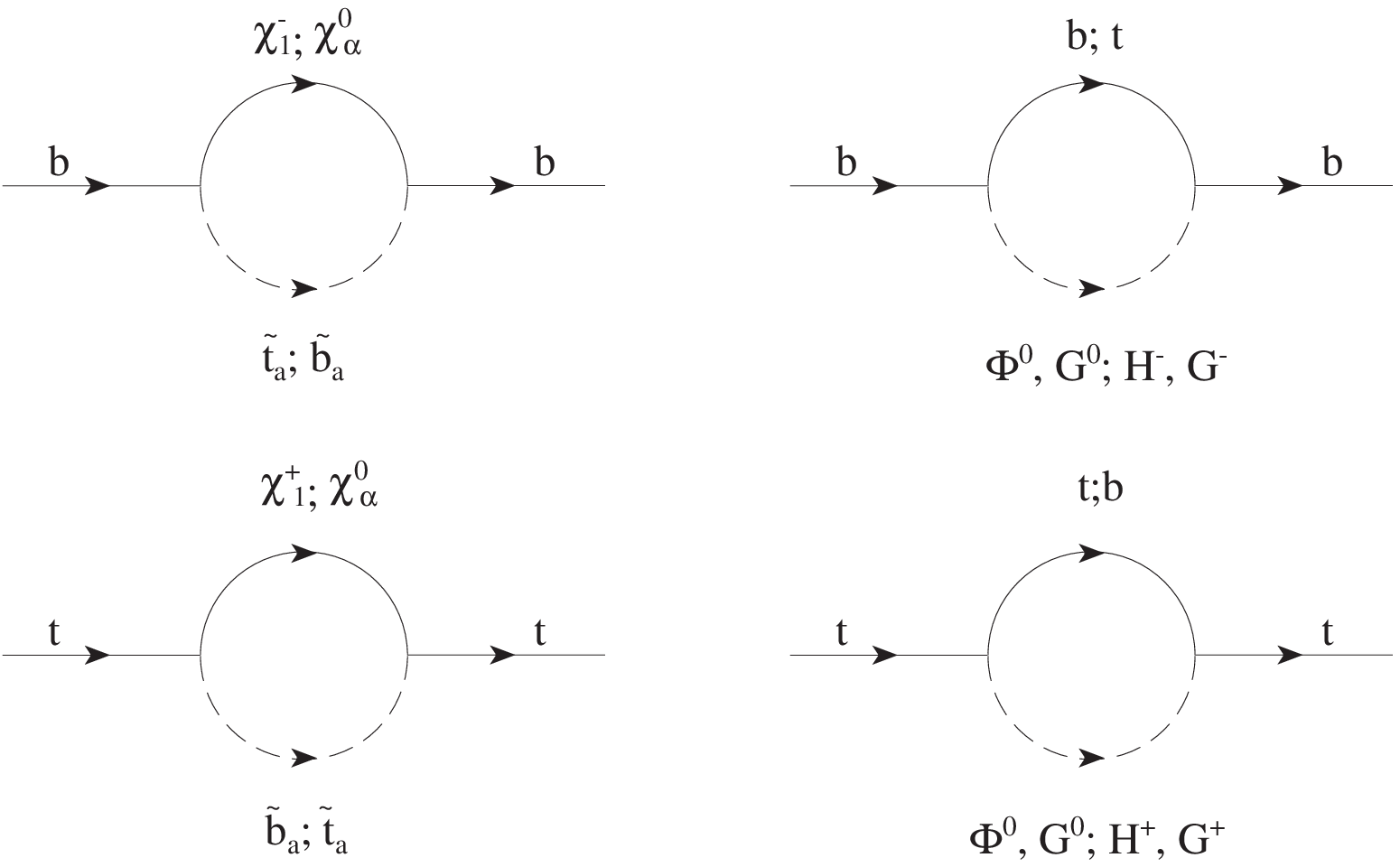,width=11cm} }\\
(a) \\ 
\mbox{\epsfig{file=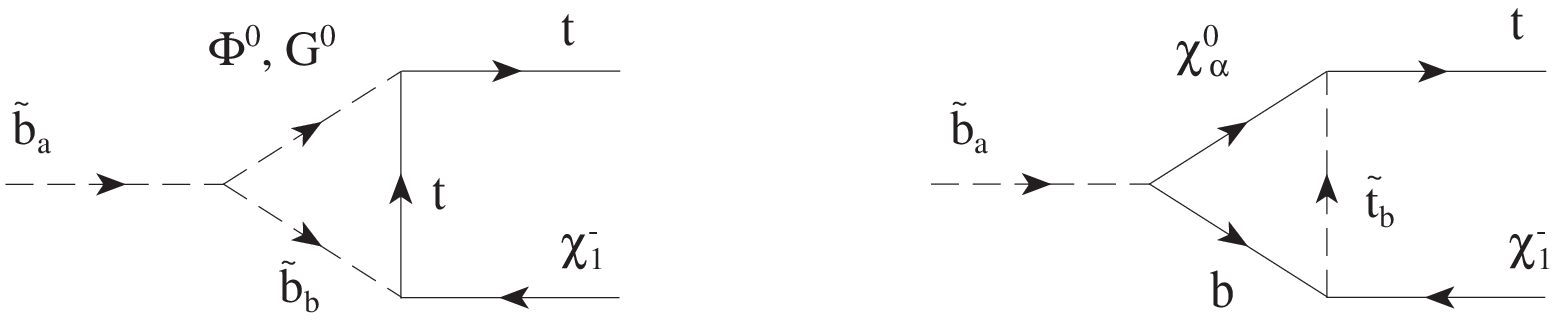,width=11cm} }\\
(b)\\ 
\end{tabular}

\caption
{Leading SUSY-EW Feynman diagrams, at one-loop order,
correcting the partial width of $\tilde{b}_a\rightarrow\chi_1^-\,t$:
{\bf (a)} Self-energy diagrams; 
{\bf (b)} Vertex contributions. Here $\Phi^0=h^0,H^0,A^0$ are the
neutral Higgs bosons of the MSSM, $H^\pm$ are the charged ones and the Goldstone
bosons are denoted $G^0,G^\pm$.
There are other possible diagrams, but
they do not contribute in the Yukawa-coupling approximation.}
\end{figure}

\begin{figure}
\centering
\begin{tabular}{c}
\mbox{\epsfig{file=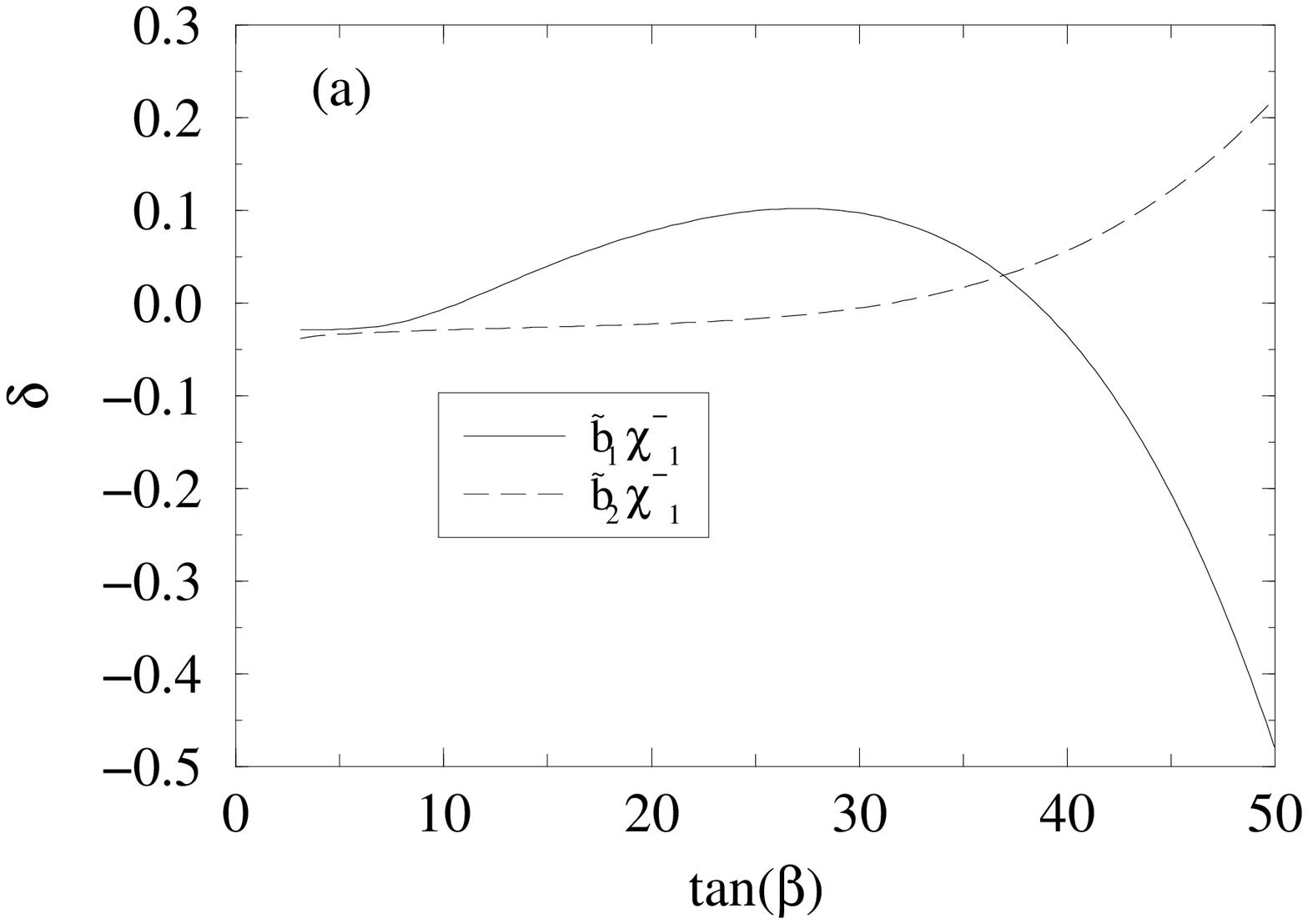,width=11cm} }\\
(a) Ê\\
\mbox{\epsfig{file=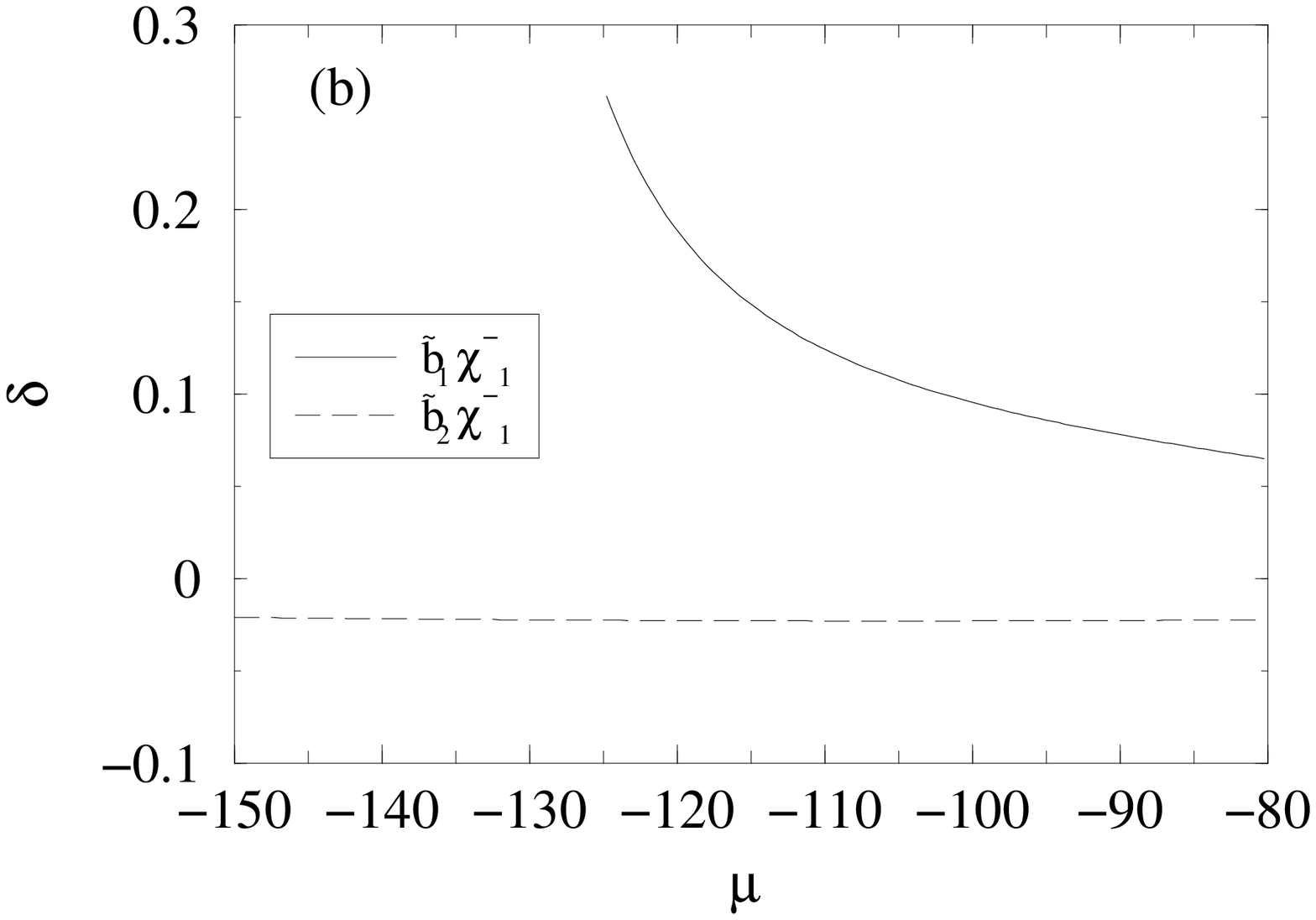,width=11cm} }\\
(b)\\ 
\end{tabular}

\caption
{{\bf (a)} The SUSY-EW correction to $\tilde{b}_a\rightarrow\chi_1^-\,t$
 as a function of 
$\tan\beta$. Fixed parameters are: $\mu=-90\,GeV$, 
$m_{\tilde{b}_1}=300\,GeV$, $m_{\tilde{b}_2}=350\,GeV$,
$m_{\tilde{t}_1}=200\,GeV$,
$\tilde{\theta}_b=\pi/10$, $\tilde{\theta}_t=\pi/5$
and $M_H=120\,GeV$; 
{\bf (b)} As in (a), but as a function of $\mu$ and fixed 
$\tan\beta=20$.}
\end{figure}

We complement the high energy analysis by considering the
possible impact of the
MSSM quantum effects on the determination of the branching ratio
$BR(t\rightarrow W^+\,b)$ from the measurement of the top quark
production cross-section at the Tevatron. The observed top quark
production cross-section is  equal to the Drell-Yan production cross-section
convoluted over the parton distributions times the
squared branching ratio. However, in the framework of the MSSM, we rather
expect a generalization of this formula in the following way (schematically),
\beqn
\sigma_{\rm obs.}&=&\int dq\,d\bar{q}\,\,\sigma (q\,\bar{q}\rightarrow t\,\bar{t})\,
\times |BR (t\rightarrow W^+\,b)|^2\nonumber\\
&+& \int dq\,d\bar{q}\,\,\sigma (q\,\bar{q}\rightarrow 
\tilde{b}_a\,\bar{\tilde{b}}_a)\,
\times |BR (\tilde{b}_a\rightarrow\chi^-_1\,t)|^2\times
|BR (t\rightarrow W^+\,b)|^2+...
\label{eq:productionMSSM}
\eeqn
We assume that gluinos are much heavier than squarks,
so that their contribution to this cross-section
through $q\,\bar{q}\rightarrow 
\tilde{g}\,\tilde{g}$ followed by
 $\tilde{g}\rightarrow t\,\tilde{t}_1$ is negligible.
From eq.(\ref{eq:productionMSSM}) we see that,
if there are alternative (non-SM) sources of top quarks subsequently
decaying into the SM final state, $W^+\,b$, then one cannot rigorously
place any stringent lower bound on
$BR (t\rightarrow W^+\,b)$ in the MSSM from the present FNAL data.
Indeed, as we have seen above we could as well have non-SM top
quark decay modes, such as e.g.
$t\rightarrow \tilde{t}_a\,\chi^0_{\alpha}$ and 
$t\rightarrow H^+\,b$ that could
serve, pictorially, as a ``sinkhole''  to compensate (at least in part) for 
the unseen source of extra top quarks produced at the Tevatron from sbottom 
pair production (Cf. eq.(\ref{eq:productionMSSM})).
However, one usually assumes that
 $BR (t\rightarrow W^+\,b)\geq 50\%$ in order to guarantee the
(purportedly) standard top quark events observed at the Tevatron. 
Thus, from these considerations it is not excluded
that the non-SM branching ratio of the top quark, 
$BR (t\rightarrow $``new''$)$,
could be as big as the SM one, i.e. $\sim 50\%$.
We stress that at present one cannot exclude eq.(\ref{eq:productionMSSM})
since the observed $t\rightarrow W^+\,b$ final state
involves missing energy, as it is also the case for the decays
comprising supersymmetric particles.
In particular, if $\tan\beta$ is large and there exists a relatively light
chargino with a non-negligible higgsino component, the alternative mechanism
suggested in eq.(\ref{eq:productionMSSM}) could be a rather efficient
non-SM source of top quarks that could compensate for the depletion in
the SM branching ratio.
From these considerations it is clear that the calculation of the 
MSSM quantum effects (Cf. Fig.\,5)
to the partial width of the decay\,\cite{DHJ,GSH} 
\beq
\tilde{b}_a\rightarrow\chi^-_i\,t\,,
\label{eq:sbottomdecay}
\eeq 
could be indispensable 
to better assess how much the determination of the SM
branching ratio $BR(t\rightarrow W^+\,b)$ is affected in the MSSM
context after plugging in the experimental number on 
the LHS of eq.(\ref{eq:productionMSSM}).

In practice the
computation of the loop contributions
to $\tilde{b}_a\rightarrow\chi^-_i\,t$
requires not only a prescription to renormalize $\tan\beta$ (we use the
same one as in the previous processes) but also a renormalization
condition for the sbottom mixing angle, $\theta_{\tilde{b}}$,
with an associated counterterm $\delta\theta_{\tilde{b}}$.
In our formalism, the 3-point Green functions explicitly incorporate 
the mixed scalar field renormalization
constants $\delta Z^{ab}$ ($a\neq b$) and are therefore renormalized also
in the $\theta_{\tilde{b}}$ parameter.
The  UV-divergent parts of the 3-point functions are cancelled against
$\delta\theta_{\tilde{b}}$ by defining the latter as follows\,\cite{GSH}:
\beq
\delta\theta_{\tilde{b}}=\frac12\,(\delta Z^{12}-\delta Z^{21})
=\frac12\,{\Sigma^{12}(m_{\tilde{b}_2}^2)+\Sigma^{12}(m_{\tilde{b}_1}^2)
\over m_{\tilde{b}_2}^2-m_{\tilde{b}_1}^2}\,.
\label{eq:rentheta}
\eeq
The QCD corrections (including the effect from gluinos) have been
studied in Ref.\cite{DHJ}. Here we concentrate on the electroweak
effects. The dominant part of them is expected to come
from the Yukawa sector, so we just consider the 
electroweak effects of ${\cal O}(h^2_t)$ 
{\it and} ${\cal O}(h^2_b)$ that emerge 
for large values of the Yukawa couplings (\ref{eq:Yukawas})
in the region of high $\tan\beta$. The leading diagrams
for the process (\ref{eq:sbottomdecay}) in this approximation are seen
in Fig.\,5.
We display the evolution of the SUSY-EW
effects as a function of $\tan\beta$ (Fig.\,6a) and of $\mu$ (Fig.\,6b).
While in Fig.\,6a $\mu=-90\,GeV$, we see in Fig.\,6b that for
$|\mu|>120\,GeV$  the corrections can be above $20\%$
for the  lightest sbottom decay ($\tilde{b}_1$). These corrections can
reach even higher values on increasing $\tan\beta$ more and more 
within the perturbative region (Cf. Fig.\,6a) and can
have either sign depending on the stop and sbottom
mixing angles\,\footnote{However, care has to be exercized in order not 
to step on prohibited colour-breaking regions of the MSSM
parameter space\,\cite{GSH}.}.
Recall that the great sensitivity
to large values of $\tan\beta$ was also the case for the MSSM
quantum corrections to $t\rightarrow H^+\,b$ (Cf. Fig.\,3b). Nevertheless,
in contrast to this latter decay, which was largely dominated by 
the bottom mass renormalization effects\,\cite{CGGJS}, the decay
(\ref{eq:sbottomdecay}) receives non-negligible contributions from
most of the diagrams in Fig.\,5. As it is manifest in Fig.\,6a, the 
(approximate) linear behaviour on $\tan\beta$ expected from bottom 
mass renormalization (Cf. Fig.\,3b) becomes
completely distorted by the rest of the contributions,
especially in the high $\tan\beta$ end.
In short, in contradistinction to the charged Higgs decay of the top
quark, the final electroweak correction to the sbottom decays
(\ref{eq:sbottomdecay})
cannot be simply ascribed 
to a single renormalization source but to the full Yukawa-coupling 
combined yield.  

\begin{figure}
\centering
\mbox{\epsfig{file=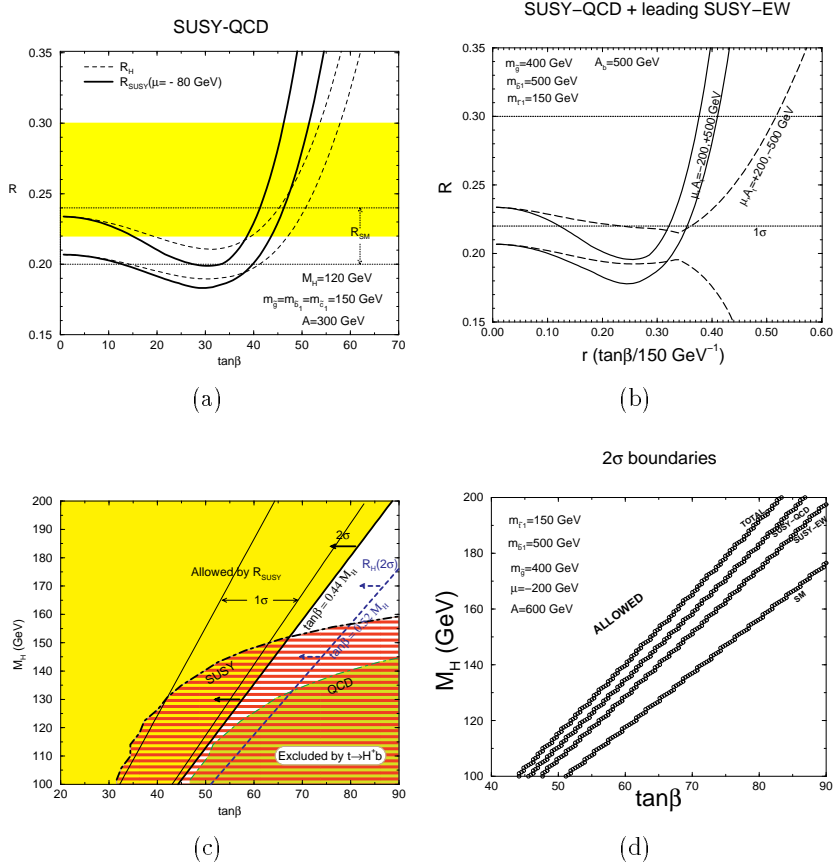,width=11cm}}
\caption
{{\bf (a)}  The SUSY-QCD corrected
ratio $R_{SUSY}$, eq.(\protect{\ref{eq:ratio}}),
as a
function of $\tan\beta$ and for fixed $\mu<0$.
Also shown are the SM result, $R_{SM}$,
(dotted band) and the Higgs-corrected
result without SUSY effects, $R_H$. 
The shaded band is the experimental measurement at the $1\,\sigma$ level;
{\bf (b)} Similar to (a) but for two cases $\mu<0$ and $\mu>0$ and including
the leading SUSY-EW effects;
{\bf (c)} The allowed region in the $(\tan\beta, M_H)$ plane for $\mu<0$;
{\bf (d)} As in (c), but showing the differences of the
$R_{SUSY}$-excluded area after including the leading SUSY-EW
effects in that ratio.}
\end{figure}

Moving now into the low energy domain, 
semileptonic $B$-meson
decays  can also reveal themselves as an invaluable probe for new physics. 
In the specific case of the inclusive semi-tauonic
$B$-meson decays, $B^-\rightarrow \tau^-\,\bar{\nu}_{\tau}\,X$
(Cf. eq.(\ref{eq:bctau})) 
one defines the following inclusive ratio 
\beq
R={\Gamma(B^-\rightarrow \tau^-\,\bar{\nu}_{\tau}\,X)\over
\Gamma (B^-\rightarrow l^-\,\bar{\nu}_{l}\,X)}\,,
\label{eq:ratio}
\eeq
where $l=e,\mu$ is a light lepton.
In Figs.7a-7d we derive the bounds obtained on the $(\tan\beta,R)$ and
$(\tan\beta, M_H)$ planes from the SUSY corrected ratio 
(\ref{eq:ratio}).
The observable $R$ is sensitive to
two basic parameters of generic
$2$HDM's, namely 
$\tan\beta$ and
the (charged) Higgs mass, $M_H\equiv M_{H^\pm}$. As a consequence, for Type II 
$2$HDM's (as in the MSSM case) the following upper bound at $1\,\sigma$
(resp. $2\,\sigma$) is claimed in the literature 
 \,\cite{QCDHaber}:
\beq
\tan\beta<0.49\ (0.52)\,\ (M_H/{\rm GeV})\,.
\label{eq:tanMH1}
\eeq 
The bound (\ref{eq:tanMH1}) also hinges on the transition from the
free quark model decay amplitude to the meson decay amplitude. In practice
this is handled within the heavy quark 
expansion formalism\,\cite{Grossman}.
Now, of course the point is whether that bound is significantly 
modified within the context of the MSSM, in particular after including 
the gluino mediated short-distance corrections. 
This SUSY-QCD analysis has been carried out in Ref.\,\cite{CJS}.
Here we improve that study by 
also including the leading SUSY-EW effects
induced by large higgsino-squark-quark Yukawa couplings, again of the type
(\ref{eq:Yukawas}). 
We find the following impact of the
SUSY effects on the physics of the semi-tauonic inclusive $B$-meson
decays within the framework of the MSSM.  For $\mu>0$, there could be 
no $\tan\beta-M_H$ bound at all (Fig.\,7b).
However, for the most
likely case $\mu<0$ (Figs.\,7a-7b and 7c-7d),
the SUSY effects further restrict the allowed region in the
$(\tan\beta,M_H)$-plane as compared to eq.(\ref{eq:tanMH1}).
The shaded region in Fig.\,7c limited by the bold solid line is allowed
at $2\,\sigma$ level by the ratio (\ref{eq:ratio}) 
after including SUSY-QCD effects. The narrow subarea between the thin
solid lines is permitted at $1\,\sigma$ level only. 
The $2\,\sigma$ region allowed without including
SUSY-QCD corrections is indicated by $R_H$, and the one excluded
by $t\rightarrow H^+\,b$ (without SUSY effects) is also shown.
Using the present day sparticle mass limits 
and the  LEP input data
on $B$-meson decays
we obtain at the $1\,\sigma$\, ($2\,\sigma)$ level\,\cite{CJS}:
\beq
\tan\beta<0.40\ (0.43)\,\ (M_H/{\rm GeV})\,.
\label{eq:tanMH2}
\eeq
While in Figs.\, 7a and 7c we consider only the SUSY-QCD corrections
to the ratio (\ref{eq:ratio}), 
in Figs.\,7b and 7d we have also included the leading SUSY-EW
effects, which amount to an additional 5-10\% strengthening of the
bound (\ref{eq:tanMH2}).

Summarizing, from the analysis of the potential SUSY quantum effects
on the semi-tauonic decays and top quark decays, a correlation seems to emerge,
namely at large $\tan\beta$ the SUSY imprint
on the corresponding low energy and high
energy  processes are both maximized. The effect is fairly large, at the level
of $20\%$ or larger. For top decays into charged Higgs, it can be of order
$50\%$. We have also found that
at present the information
on $(\tan\beta,M_H)$ as collected from $B$-meson decays
is more restrictive
than the one from top quark decays in certain regions
of parameter space, specifically those which are phase-space inaccessible
to top quark decay. However, in the phase-space accessible region,
the data from top quark physics is more restrictive. Clearly,
knowledge from both low energy and high energy data can be very useful
to better pinpoint in the future the physical boundaries 
of the MSSM parameter space.
Alternatively, if the two approaches would converge to 
a given portion of that parameter
space, one could claim strong indirect evidence of SUSY. 
 
Finally, we have also shown that at large $\tan\beta$ there are other 
SUSY decay processes and production mechanisms that could be
severely influenced by the MSSM quantum corrections. These non-SM effects, 
in turn, could modify the determination of much more ``down-to-earth''
observables such as the branching ratio of the standard
top quark decay. The unexpected and very much rewarding result of this
is that although the SM decay
$t\rightarrow W^+\,b$ proves to be by itself rather insensitive
to SUSY loop effects (Cf. Fig.\,3a), the chain of 
intermediate mechanisms leading to top quark production (with the
inclusion of potential SUSY sources of top quarks)
could be, instead, much more sensitive to them (Cf. Fig.\,6).
At the end of the day it turns out that, despite the absence
of noticeable SUSY quantum effects
on the standard decay of the top quark,
the determination of 
$BR(t\rightarrow W^+\,b)$ from the observed cross-section
could effectively incorporate a significant quantum SUSY signature!.

{\bf Acknowledgements}:
The author is grateful to the organizers
of WHEPP-5 for finantial support and for the warm hospitality offered to him
in IUCAA, Pune Univ., where the workshop was held. 
Discussions with D. P. Roy, P. Roy and M. Guchait are acknowledged.
He is also thankful to Toni Coarasa and Jaume Guasch for their
help in the preparation of this manuscript.
This work has been partially supported by CICYT under
project No. AEN93-0474.
\noindent



\baselineskip=5.5mm
\vspace{1cm}
\begin{thebibliography}{9999}
\bibitem{Gunion}
J. F. Gunion,\, talk given at the {\it Int. Workshop on
Quantum Effects in the MSSM}, Barcelona, September 9-13 
(1997)  [hep-ph/9801417], to appear in  the proceedings, 
World Scientific 1998, Ed. J. Sol\`a.
 \bibitem{WdeBoer}
W. Hollik, talk given at the {\it Int. Workshop on Quantum Effects
in the MSSM}, Barcelona, September 9-13 (1997) [hep-ph/9711489],
to appear in  the proceedings, World Scientific 1998, Ed. J. Sol\`a;
W. de Boer, A. Dabelstein, W. Hollik, W. M\"osle, U. Schwickerath,
{\it Z. Phys.} {\bf C 75} (1997) 627.
\bibitem{Peskin}
M. Peskin,\, lectures given at the {\it European School of
High-Energy Physics} [hep-ph/9705479]. 
\bibitem{CLEO}
M.S. Alam et al. (CLEO Collab.)\, {\it Phys. Rev. Lett.}\,
{\bf 74} (1995) 2885; S. Glenn (for the CLEO Collab.), talk at the
Meeting of the American Physics Society, Ohio, March 1998. 
\bibitem{Hunter}
J.F. Gunion, H.E. Haber, G.L. Kane, S. Dawson,\, {\it The Higgs Hunters'
Guide} (Addison-Wesley, Menlo-Park, 1990).
\bibitem{Ciuchini}
M. Ciuchini, G. Degrassi, P. Gambino, G.F. Giudice,
[hep-ph/9710335]; F. Borzumati, C. Greub, [hep-ph/9802391].
\bibitem{ALEPH}
P.M. Kluit,\,talk at the Int. Europhysics Conference, 1997.
\bibitem{Ciuchini2}
M. Ciuchini, G. Degrassi, P. Gambino, G.F. Giudice,\,
[hep-ph/9806308].
\bibitem{CGGJS}
J. A. Coarasa, D. Garcia, J. Guasch, R.A. Jim\'enez, 
J. Sol\`a,\, {\it Eur. Phys. J.}\, {\bf C2} (1998) 373;
J. Guasch, R.A. Jim\'enez, J. Sol\`a,\, {\it Phys. Lett.}\,
{\bf B 360} (1995) 47. 
\bibitem{Grossman}
Y. Grossman, Z. Ligeti\, {\it Phys. Lett.}\, {\bf B 332} (1994) 373.
\bibitem{Guasch1}
J. Guasch, J. Sol\`a,\,
{\it Z. Phys.}\, {\bf C 74} (1997) 337.
\bibitem{GJSH}
D. Garcia, W. Hollik, R.A. Jim\'enez, J. Sol\`a,\, {\it Nucl. Phys.}
 {\bf B 427} (1994) 53;
A. Dabelstein, W. Hollik, R.A. Jim\'enez, C. J\"unger,
J. Sol\`a,\, {\it Nucl. Phys.}\, {\bf B 456} (1995) 75.
\bibitem{CDF}
B. Bevensee, talk given at the {\it Int. Workshop on Quantum Effects
in the MSSM}, Barcelona, September 9-13 (1997), to appear in 
the proceedings, World Scientific 1998, Ed. J. Sol\`a;
F. Abe {\it et al.} (CDF Collab.),\, {\it Phys. Rev. Lett.}\, {\bf 79}
(1997) 357.
\bibitem{SO10}
M. Carena, S. Pokorski, 
C.E.M. Wagner,\,{\it Nucl. Phys.} {\bf B 426} (1994) 269;
L.J. Hall, R. Rattazzi, U. Sarid,\, {\it Phys. Rev.} {\bf D 50}
 (1994) 7048.
\bibitem{Guasch4}
J. Guasch, J. Sol\`a,\, {\it Phys. Lett.}\, {\bf B 416} (1998) 353.  
\bibitem{DPRoy}
 M. Guchait, D.P. Roy,\, {\it Phys. Rev.}\, {\bf D 55} (1997) 7263.
\bibitem{CGGJS2}
J. A. Coarasa, D. Garcia, J. Guasch, R.A. Jim\'enez, 
J. Sol\`a, \, {\it Phys. Lett.}\, {\bf B 425} (1998) 329.
\bibitem{Guchait1}
M. Drees, M. Guchait, P. Roy,\,{\it Phys. Rev. Lett.}\, {\bf 80} (1998) 2047.
\bibitem{Coarasa}
J.A. Coarasa, R.A. Jim\'enez, J. Sol\`a,
{\it Phys. Lett.}\,  {\bf B 389} (1996)  312.
\bibitem{DHJ} A. Djouadi, W. Hollik, C. J{\"u}nger,\, 
{\it Phys. Rev.}\, {\bf D 55} (1997) 6975;
S. Kraml, H. Eberl, A. Bartl, W. Majerotto,
W. Porod,\, {\it Phys. Lett.} {\bf B 386} (1996) 175.
\bibitem{GSH}
J. Guasch, W. Hollik, J. Sol\`a,\, [hep-ph/9802329].
\bibitem{QCDHaber}
Y. Grossman, H.E. Haber, Y. Nir,\, {\it Phys. Lett.}\, {\bf B 357} (1995) 630.
\bibitem{CJS}
J.A. Coarasa, R.A. Jim\'enez, J. Sol\`a,
{\it Phys. Lett.}\,  {\bf B 406} (1997)  337.


\end{thebibliography}
\end{document}